%% file: ms.tex
\documentclass[sigconf]{acmart}
\pdfoutput=1

\usepackage{graphicx} %
\usepackage{natbib}  %
\usepackage{caption} %
\usepackage{algorithm}
\usepackage{listings}
\usepackage{algorithmic}
\usepackage{amsmath}
\usepackage{booktabs}
\usepackage{multirow}
\usepackage[outdir=./]{epstopdf}
\usepackage{enumitem}
\usepackage{caption}
\usepackage{subcaption}
\usepackage{newfloat}

\AtBeginDocument{%
  \providecommand\BibTeX{{%
    \normalfont B\kern-0.5em{\scshape i\kern-0.25em b}\kern-0.8em\TeX}}}

\copyrightyear{2022} 
\acmYear{2022} 
\setcopyright{rightsretained} 
\acmConference[MM '22]{Proceedings of the 30th ACM International Conference on Multimedia}{October 10--14, 2022}{Lisboa, Portugal}
\acmBooktitle{Proceedings of the 30th ACM International Conference on Multimedia (MM '22), Oct. 10--14, 2022, Lisboa, Portugal}
\acmISBN{978-1-4503-9203-7/22/10}
\acmDOI{10.1145/3503161.3548428}

\usepackage{etoolbox}
\makeatletter
\patchcmd{\maketitle}{\@copyrightpermission}{
   \begin{minipage}{0.3\columnwidth}
     \href{https://creativecommons.org/licenses/by/4.0/}{\includegraphics[width=0.90\textwidth]{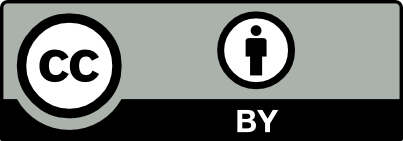}}
   \end{minipage}\hfill
   \begin{minipage}{0.7\columnwidth}
     \href{https://creativecommons.org/licenses/by/4.0/}{This work is licensed under a Creative Commons Attribution International 4.0 License.}
   \end{minipage}
  
   \vspace{5pt}
}{}{}

\makeatother

\begin{document}

\title{DVR: Micro-Video Recommendation Optimizing Watch-Time-Gain under Duration Bias}

\author{Yu Zheng}
\orcid{0000-0002-1837-6730}
\affiliation{
  \institution{Department of Electronic Engineering, Tsinghua University}
  \city{Beijing}
  \country{China}
}

\author{Chen Gao}
\authornote{Corresponding author (chgao96@gmail.com).}
\affiliation{%
  \institution{Department of Electronic Engineering, Tsinghua University}
  \city{Beijing}
  \country{China}
}

\author{Jingtao Ding}
\affiliation{
  \institution{Department of Electronic Engineering, Tsinghua University}
  \city{Beijing}
  \country{China}
}

\author{Lingling Yi}
\affiliation{
  \institution{WeChat Technical Architecture Department, Tencent Inc.}
  \city{Shenzhen}
  \country{China}
}

\author{Depeng Jin}
\affiliation{
  \institution{Department of Electronic Engineering, Tsinghua University}
  \city{Beijing}
  \country{China}
}

\author{Yong Li}
\affiliation{
  \institution{Department of Electronic Engineering, Tsinghua University}
  \city{Beijing}
  \country{China}
}

\author{Meng Wang}
\affiliation{
  \institution{School of Computer Science and Information Engineering, Hefei University of Technology}
  \city{Hefei}
  \country{China}
}

\renewcommand{\shortauthors}{Yu Zheng et al.}

\begin{abstract}
Recommender systems are prone to be misled by biases in the data.
Models trained with biased data fail to capture the real interests of users, thus it is critical to alleviate the impact of bias to achieve unbiased recommendation.
In this work, we focus on an essential bias in micro-video recommendation, duration bias.
Specifically, existing micro-video recommender systems usually consider \textit{watch time} as the most critical metric, which measures how long a user watches a video.
Since videos with longer duration tend to have longer watch time, there exists a kind of \textit{duration bias}, making longer videos tend to be recommended more against short videos.
In this paper, we empirically show that commonly-used metrics are vulnerable to duration bias, making them NOT suitable for evaluating micro-video recommendation.
To address it, we further propose an unbiased evaluation metric, called \textbf{WTG} (short for \textit{\textbf{W}atch \textbf{T}ime \textbf{G}ain}).
Empirical results reveal that WTG can alleviate duration bias and better measure recommendation performance.
Moreover, we design a simple yet effective model named \textbf{DVR} (short for \textit{\textbf{D}ebiased \textbf{V}ideo \textbf{R}ecommendation}) that can provide unbiased recommendation of micro-videos with varying duration, and learn unbiased user preferences via adversarial learning.
Extensive experiments based on two real-world datasets demonstrate that DVR successfully eliminates duration bias and significantly improves recommendation performance with over 30\% relative progress.
Codes and datasets are released at \url{https://github.com/tsinghua-fib-lab/WTG-DVR}.
\end{abstract}



\maketitle

\input{1.introduction.tex}
\input{2.metrics.tex}

\input{3.model.tex}
\input{4.experiments.tex}

\input{5.related.tex}

\input{6.conclusion.tex}

\begin{acks}
This work is supported in part by The National Key Research and Development Program of China under grant 2020AAA0106000. This work is also supported in part by the National Natural Science Foundation of China under U1936217, 61971267, 61972223, U20B2060.
\end{acks}

\clearpage
\balance
\bibliographystyle{ACM-Reference-Format}
\bibliography{bibliography}

\clearpage
\nobalance
\appendix
\input{7.appendix}

\end{document}

%% file: 1.introduction.tex
\section{Introduction}\label{sec::intro}
Today's micro-video platforms, such as TikTok\footnote{\url{https://www.tiktok.com/}}, have been taking the majority of Internet traffic. 
With millions of micro-videos uploaded per day, 
recommender systems have become the fundamental channel that users access micro-videos~\cite{davidson2010youtube,covington2016deep,wei2019mmgcn,chen2019top,beutel2018latent,li2019routing,chen2018temporal,gao2017unified,wei2019personalized,gao2021graph_survey,zhang2021group}.
Existing approaches usually consider watch time as a critical index of user satisfaction and activeness, thus recommend micro-videos with higher estimated watch time \cite{covington2016deep}.
Specifically, recommender systems take rich features like user profiles and video attributes as input, 
and predict watch time with a parametric model such as deep neural networks \cite{covington2016deep}.
Micro-videos with longer predicted watch time are ranked higher and recommended to the users.
However, longer watch time does not necessarily indicate that the user is more interested in the micro-video, since watch time is highly correlated with the duration of the video.
Such duration bias makes it challenging to evaluate the performance and learn user preferences for micro-video recommendation.

\begin{figure}[t]
    \centering
    \includegraphics[width=\linewidth]{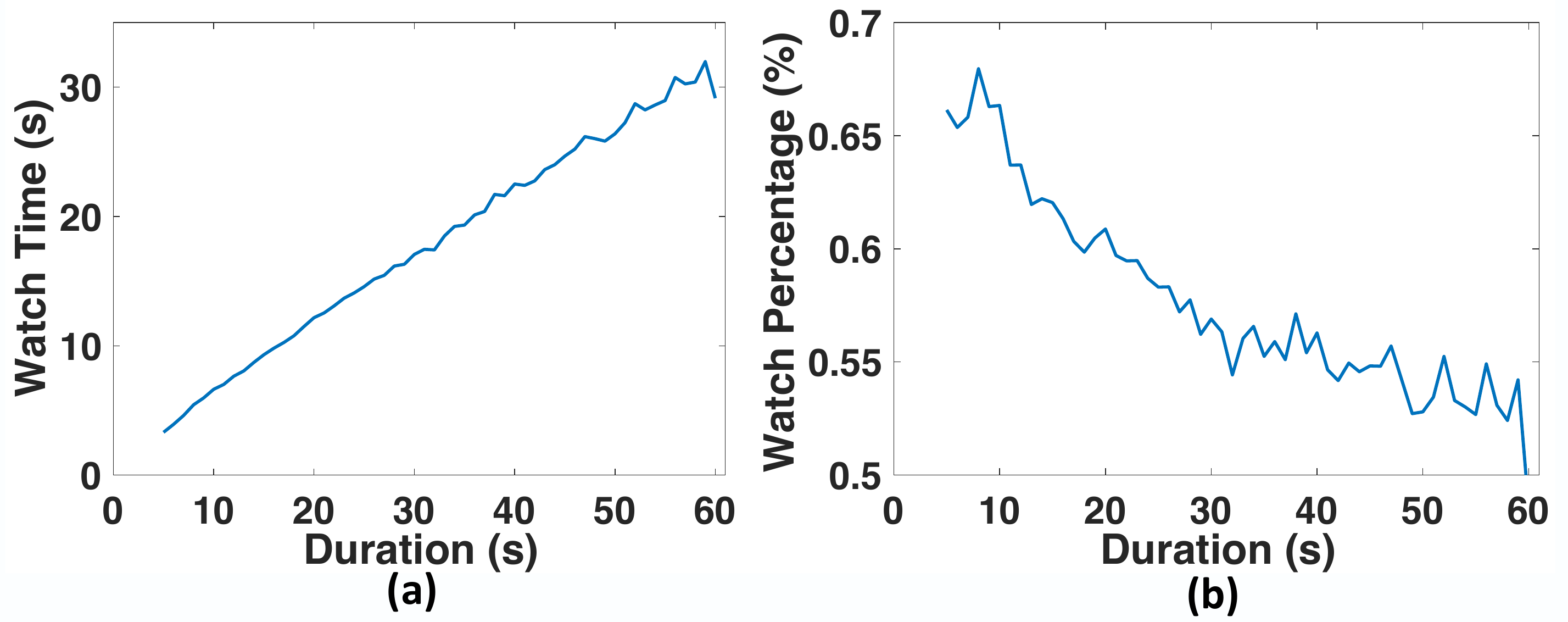}
    \vspace{-20px}
    \caption{Duration bias of micro-videos with different duration. (a) Mean watch time (b) Mean watch percentage .}
    \vspace{-20px}
    \label{fig::duration_bias}
\end{figure}

The duration bias hidden in user-video interaction data means that micro-videos with longer duration tend to have longer watch time, since users usually decide whether to continue watching or switch to the next one until watching a certain fraction of the micro-video.
Here the \textit{duration} is defined as \textit{the total length of a micro-video}.
Figure \ref{fig::duration_bias} (a) shows the duration and average watch time of a real-world micro-video dataset, Wechat Channels (details of the dataset will be introduced later in Section \ref{sec::dataset}).
The watch time grows as duration increases, which demonstrates the existence of duration bias.
To address it, an intuitive solution is to use ``watch percentage'' instead of watch time.
Unfortunately, micro-videos of short duration tends to have larger watch percentage, which means that duration bias still exists but in the opposite direction, illustrated in Figure \ref{fig::duration_bias} (b).
It is worthwhile to notice that Figure \ref{fig::duration_bias} (a) and Figure~\ref{fig::duration_bias} (b) actually describe the same phenomenon and Figure \ref{fig::duration_bias} (b) can be obtained by normalizing Figure \ref{fig::duration_bias} (a) with the video duration (x-axis).
In addition, our finding of duration bias is in line with related literature \cite{wu2018beyond} on measuring user engagement on online videos.
In the following, we will elaborate on how the duration bias leads to two main undesired consequences: \textit{inaccurate} recommendation and \textit{unfair} recommendation.

\textbf{Inaccurate recommendation caused by duration bias.}
Unlike traditional scenarios that deal with discrete user feedback, such as rating prediction~\cite{koren2009matrix,koren2008factorization,koren2009collaborative}, implicit collaborative filtering (CF)~\cite{hu2008collaborative,rendle2012bpr,he2017ncf} and click-through rate (CTR) prediction~\cite{zhou2018deep,song2019autoint,liu2020autofis}, user engagement towards videos is mainly reflected by the watch time, which is continuous \cite{salganik2006experimental,krumme2012quantifying}.
Specifically, a user tends to continue watching if he/she is interested in the current video, and otherwise, he/she may switch to the next one.
In other words, the continuous value, watch time, serves as a substantial indicator of user preference.
However, caused by duration bias, longer watch time does not necessarily mean that users are more interested, which we have shown in Figure \ref{fig::duration_bias}.
As a consequence, a recommendation model can be easily misled by the duration bias, and recommend too many micro-videos that do not match user preference but with long duration.
It is worthwhile to notice that micro-video platforms like TikTok insert advertisements between different micro-videos, thus simply recommending long micro-videos will NOT bring higher advertising revenue, which is different from platforms like YouTube that insert advertisements inside the videos.

\begin{figure}[t]
    \centering
    \includegraphics[width=\linewidth]{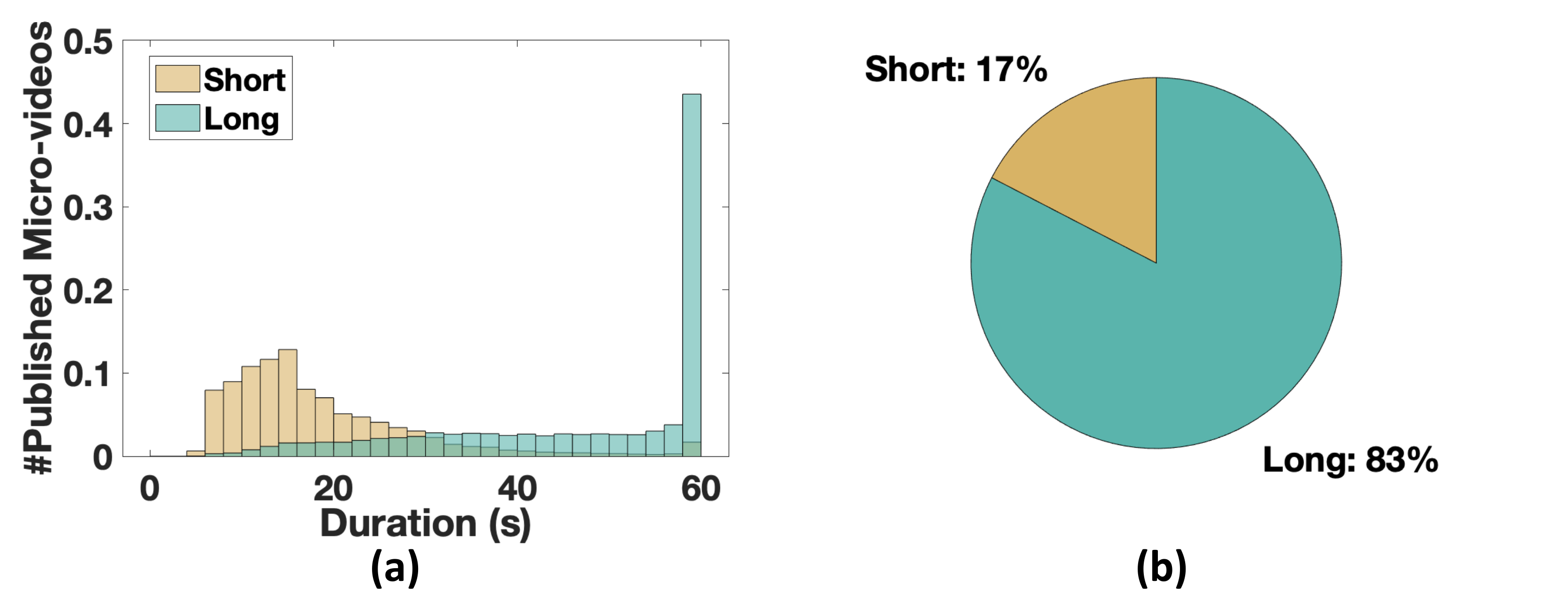}
    \vspace{-20px}
    \caption{(a) The distribution of uploaded micro-videos for two user groups; (b) The recommendation traffic received by the two user groups.}
    \vspace{-20px}
    \label{fig::unfair}
\end{figure}

\textbf{Unfair recommendation caused by duration bias.}
On the other hand, different users upload micro-videos of different duration, ranging from a few seconds like short funny videos to longer ones of a few minutes like VLogs.
As we have mentioned, such duration bias makes longer videos more likely to be recommended than shorter videos, which favors long video publishers and is unfair for short video publishers.
To show this point, we compute the average duration of the uploaded micro-videos for each user on the above Wechat Channels dataset, and separate all the micro-video producers from the middle into two groups, which are long micro-video producers and short micro-video producers.
Figure \ref{fig::unfair} (a) illustrates the quite different distributions of the published micro-videos with respect to the duration of the two user groups.
Then we implement the famous Factorization Machine (FM) model~\cite{rendle2010factorization}, and show the recommendation chances received by the two groups in Figure \ref{fig::unfair} (b).
We can observe that ``long'' group receives much more recommendation chances with over 80\% than ``short'' group with less than 20\%, although the two groups have the same number of users.
Such results show that recommendation based on watch time lead to unfairness for shorter-video publishers.

In this paper, we take the first step to eliminate duration bias
for micro-video recommendation.
Specifically, in order to reduce the impact of duration bias on both evaluation and learning of micro-video recommendation, we investigate two research questions.
\begin{itemize}[leftmargin=*]

\item \textbf{RQ1: How to measure users' watch time towards micro-videos in an unbiased way?} 
Since traditional metrics, such as total watch time and total watch percentage of the top $k$ recommended micro-videos, all suffer from duration bias, which favor either long or short micro-videos, it is crucial to define an unbiased metric that does not favor either side.

\item \textbf{RQ2: How to learn unbiased user preferences on micro-videos of different duration and provide accurate recommendation?}
Existing recommendation approaches are vulnerable to the duration bias since the duration of micro-videos is a strong feature when predicting watch time.
Therefore, designing recommender systems that are free from duration bias is useful to capture users' real interests in micro-videos.

\end{itemize}

Alleviating duration bias for micro-video recommendation is largely unexplored, and we face two main challenges. 
\textbf{First}, micro-videos of different duration can not be compared directly.
The final watch time of a micro-video is determined by both user preference and video duration.
Therefore, watch time and video duration need to be compared jointly to evaluate the performance with respect to user preference.
\textbf{Second}, since the structural differences between recommendation models vary widely, the bias alleviation design is supposed to be general and model-agnostic.
In other words, it needs to be compatible with any recommendation model that ranks micro-videos according to rich input features.

For the first research question, we propose an unbiased evaluation metric \textbf{W}atch \textbf{T}ime \textbf{G}ain (WTG), which measures a user's relative engagement on a video against the average engagement of all users on videos with the same duration-level.
The proposed metric overcomes the influence of video duration, and videos of different duration are forced to be \textit{flattened} equally, and they are comparable with each other, which addresses the first challenge.
Meanwhile, to emphasize the order of recommended micro-videos, \textit{i.e.} micro-videos of larger WTG are best to rank higher in the recommendation list, we further propose a \textbf{D}iscounted \textbf{C}umulative version of WTG (DCWTG) inspired by the widely adopted NDCG metric in recommendation literature \cite{he2017ncf,zheng2021disentangling,chang2021sequential}.
The proposed WTG and DCWTG provide unbiased evaluation protocols for micro-video recommendation.
For the second research question, we further propose a framework named \textbf{D}ebiased \textbf{V}ideo \textbf{R}ecommendation (DVR), which can learn user preference with simple and effective strategies to remove the duration bias and facilitate accurate recommendation.
The proposed DVR framework adds an adversarial layer on the predicted value of existing recommendation models, and it does not have any preset requirements for the structure of the backbone models.
Therefore, it can be combined with any off-the-shelf recommender systems, which addresses the second challenge.

We conduct experiments on two real-world datasets collected from the largest micro-video platforms in China.
Specifically, we perform a large-scale analysis to investigate the impact of duration bias and the shortcomings of existing metrics of micro-video recommendation.
In addition, we demonstrate that the proposed metric WTG can measure users' watch time on micro-videos in an unbiased way which does not favor long or short videos.
Furthermore, we show that WTG can help construct unbiased recommender systems.
We combine DVR with various backbone models, and experimental results show that DVR can improve state-of-the-art recommendation approaches with over 30\% relative progress.

The main contributions of the paper are summarized as follows:
\begin{itemize}[leftmargin=*]
    \item We take the pioneering step to alleviate duration bias for micro-video recommendation. We conduct a large-scale analysis to show how duration bias leads to inaccurate recommendation.
    
    \item We propose a new metric, WTG, to achieve unbiased measurement of users' watch time on micro-videos, which eliminates duration bias.
    We further propose a novel model DVR to learn unbiased user preference on micro-videos of different duration.
    
    \item Extensive experiments on two real-world datasets show that our proposed metrics and DVR model successfully achieve unbiased recommendation of micro-videos.
\end{itemize}

%% file: 2.metrics.tex
\section{Data Analysis}\label{sec::analysis}
As introduced previously in Figure \ref{fig::duration_bias}, micro-videos with longer duration tend to have longer watch time.
In this section, we further investigate the impact of such duration bias on recommendation models.
We conduct analysis on the same dataset as Figure \ref{fig::duration_bias} and \ref{fig::unfair}, and the details of the adopted dataset will be introduced in Section \ref{tab::dataset}.
Specifically, we select representative and state-of-the-art recommendation models that aim to predict watch time, following the common paradigm in existing recommender systems.
Then we demonstrate that these models are influenced by the duration bias, which makes them recommend too many micro-videos that do not match user preference but with long duration.
Meanwhile, we also compare these methods with several intuitive and trivial approaches, such as always recommending long videos, and reveal the shortcomings of existing metrics.

\begin{figure}[t]
    \centering
    \includegraphics[width=\linewidth]{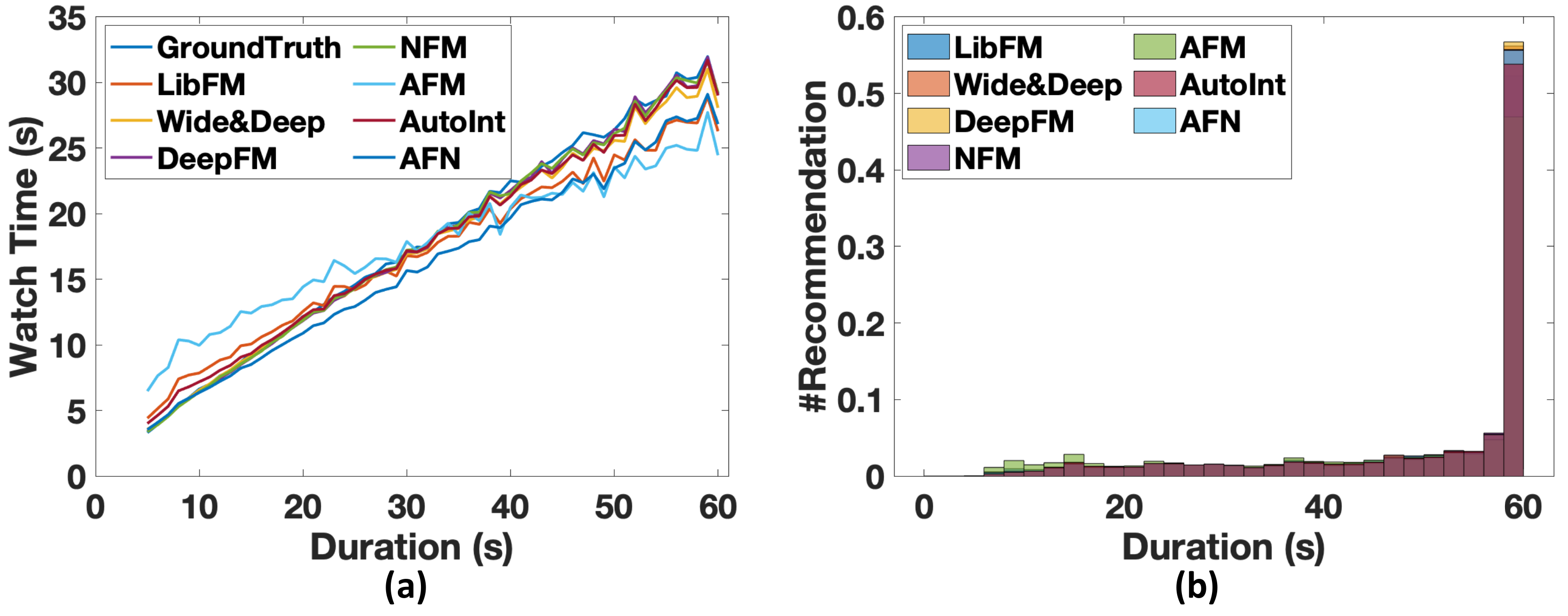}
    \vspace{-20px}
    \caption{(a) Average predicted watch time of different models. (b) Histogram of recommended micro-videos with different duration of different models.}
    \vspace{-20px}
    \label{fig::wechat_pred}
\end{figure}

\noindent\textbf{Distribution shift of recommendation results.}
In order to study the influence of duration bias on micro-video recommendation, we use recommendation models to predict the watch time of micro-videos and analyze the results.
We adopt classical algorithms including LibFM \cite{rendle2010factorization}, Wide\&Deep \cite{cheng2016wide}, DeepFM~\cite{guo2017deepfm}, NFM \cite{he2017nfm} and AFM~\cite{xiao2017attentional}, as well as state-of-the-art approaches including AutoInt \cite{song2019autoint} and AFN \cite{cheng2020adaptive}.
We first calculate the average predicted watch time of micro-videos with different duration.
We discretize duration into equally sized bins with a width of 1 second.
Figure \ref{fig::wechat_pred} (a) shows the predicted and groundtruth watch time of micro-videos in different bins.
We can observe that models tend to \textit{amplify} duration bias in the data.
Specifically, we can find in \ref{fig::wechat_pred} (a) that the slope of curves of all models are much higher than the groundtruth curve (blue). 
In other words, the predicted watch time of long (short) videos is much longer (shorter) than it supposed to be.
Such bias amplification of recommendation models is damaging to user experience, since they recommend too many micro-videos with long duration.
To illustrate this point, we select the top $k$ recommended micro-videos of all models and plot the histogram of their duration in Figure \ref{fig::wechat_pred} (b).
We can discover that recommended micro-videos concentrate on the long duration side, while all the models almost do not recommend any short micro-videos.
We now show that such biased recommendation is not only inaccurate but also unfair.

\begin{itemize}[leftmargin=*]
\item \textit{Inaccurate recommendation due to the distribution shift.}
Simply recommending micro-videos with a long duration can not meet users' needs which leads to inaccurate recommendation, since there are a large amount of \textit{bad} cases of long micro-videos, \textit{i.e.} users may quickly find that they have no interest in the long micro-video and switch to the next one.
Therefore, we calculate the number of bad cases for these recommendation models.
Specifically, we define bad cases as the recommended micro-videos with groundtruth watch time lower than 2 seconds.
In order to reveal the shortcomings of using watch time for micro-video recommendation, we add two trivial models \textbf{LongRec} and \textbf{RandomRec},
where LongRec model ranks the micro-videos directly according to the duration thus long micro-videos rank higher, and RandomRec model just randomly shuffles the micro-videos to provide a recommendation list.
Table \ref{tab::analysis} shows the results of all the above models with respect to Mean Absolute Error (\textbf{MAE}), Root Mean Squared Error (\textbf{RMSE}), total Watch Time of top $k$ videos (\textbf{WatchTime@k}) and the number of bad cases (\textbf{\#BC}).
We have two important observations.
First, both classical models and state-of-the-art models achieve comparable WatchTime@k with the trivial Long model.
In other words, although these models have a strong capacity with thousands of learnable parameters, they fail to learn much more than the duration bias.
Second, bad cases generated by these models are only slightly less than the \textit{Random} model, which means that it is not a reasonable choice to recommend blindly according to the predicted watch time.
Such many bad cases indicate that the duration bias results in low recommendation accuracy.
Therefore, it is crucial to define an unbiased metric to measure user engagement towards micro-videos.
In addition, an unbiased metric can also facilitate user preference learning to make it free from the influence of video duration.

    \item \textit{Unfair recommendation due to the distribution shift.} We compare the recommendation traffic received by long and short micro-video publishers.
Specifically, we use the above well-trained models, and rank the micro-videos according to the estimated watch time.
Then for each user, we recommend $k$ micro-videos with the highest estimated watch time.
We vary the value of $k$, and Figure \ref{fig::traffic_k} illustrates the recommendation traffic received by users who mainly produce long or short micro-videos.
We can observe that short micro-video publishers hardly receive any recommendation when $k$ is small, and they only receive less than 20\% of recommendation traffic even with a large enough value of $k$.
Meanwhile, long micro-video publishers obtain much more recommendation for their videos than short micro-video publishers.
Comparison of the recommendation traffic verifies that recommending micro-videos based on predicted watch time leads to serious unfairness for different micro-video publishers.

\end{itemize}

\begin{table}[tb]
\centering
\small
\caption{The impact of duration bias (larger WatchTime@k and smaller MAE, RMSE, \#BC means better performance). We can observe that personalized models are even as poor as non-personalized RandomRec or LongRec.}
\vspace{-10px}
\label{tab::analysis}
\begin{tabular}{c|cccc}
\toprule
\bf{Model}      & \bf{MAE}     & \bf{RMSE}    & \bf{WatchTime@k} & \bf{\#BC} \\
\midrule
RandomRec     & 12.18 & 18.21 & 117.41    &    3850  \\
LongRec       & 6.30  & 12.50 & 202.92    &   3679 \\ \hline
LibFM      & 5.48  & 8.13  & 204.72    &   3560   \\
Wide\&Deep & 5.26  & 7.85  & 205.75    &   3558   \\
DeepFM     & 5.29  & 7.85  & 205.83    &   3553   \\
NFM        & 5.23  & 7.82  & 206.08    &   3550   \\
AFM        & 6.67  & 9.86  & 158.48    &   3515   \\
AutoInt    & 5.23  & 7.86  & 205.77    &   3568      \\
AFN        & 5.72  & 8.44  & 201.56    &   3536   \\
\bottomrule
\end{tabular}
\vspace{-10px}
\end{table}

In short, we have the following observations from data analysis.
\begin{itemize}[leftmargin=*]
    \item Duration bias is amplified by recommendation models, leading to unbalanced recommendation results: the model recommends much more long micro-videos than short ones.
    
    \item Such unbalance leads to inaccurate recommendation and a large number of bad recommendation cases.
    
    \item The unbalance leads to unfairness, favoring long micro-video producers, which is unfair for short micro-video producers.

\end{itemize}

%% file: 3.model.tex
\section{Method}\label{sec::method}
To alleviate the duration bias that leads to inaccurate and unfair recommendation, we propose a new unbiased metric of watch time, WTG, and an unbiased recommendation model, DVR.

\subsection{WTG: An Unbiased Metric of Watch Time}
\textbf{Watch Time Conditioned on Duration.} Based on the above analysis, we can conclude that watch time can not be directly used as an indicator for user engagement/preference since watch time is to a great extent dominated by the duration bias.
However, if we condition on the value of micro-video duration, watch time can be regarded as a reasonable metric on whether the micro-video matches the user's preference \cite{wu2018beyond}.
For example, a user $u$ may watch 50 seconds of a 60s-duration micro-video $v_i$, and only watch 5 seconds of another micro-video $v_j$ with the same duration of 60 seconds.
Then we can confidently infer that the user $u$ prefers micro-video $v_i$, while he/she might have little interest in micro-video $v_j$.
In other words, the watch time of a micro-video can be used as a metric only when it is compared with other data points of micro-videos with similar duration.
Therefore, we define a new metric called Watch Time Gain \textbf{(WTG)}, which measures the relative user engagement on a micro-video compared with the average engagement of all users on micro-videos with a similar duration.
Specifically, we first divide all the micro-videos into equally wide bins according to their duration, and each micro-video can be mapped to its corresponding duration bin as follows,
\begin{align}
    &\mathbf{B} = [b_1, \cdots, b_m], \\
    &B(v) = f_b(d_v),
\end{align}
where $m$ is the number of bins, $d_v$ is the duration of micro-video $v$, and $f_b$ is a function mapping a duration to the corresponding bin.

\begin{figure}[t]
    \centering
    \includegraphics[width=\linewidth]{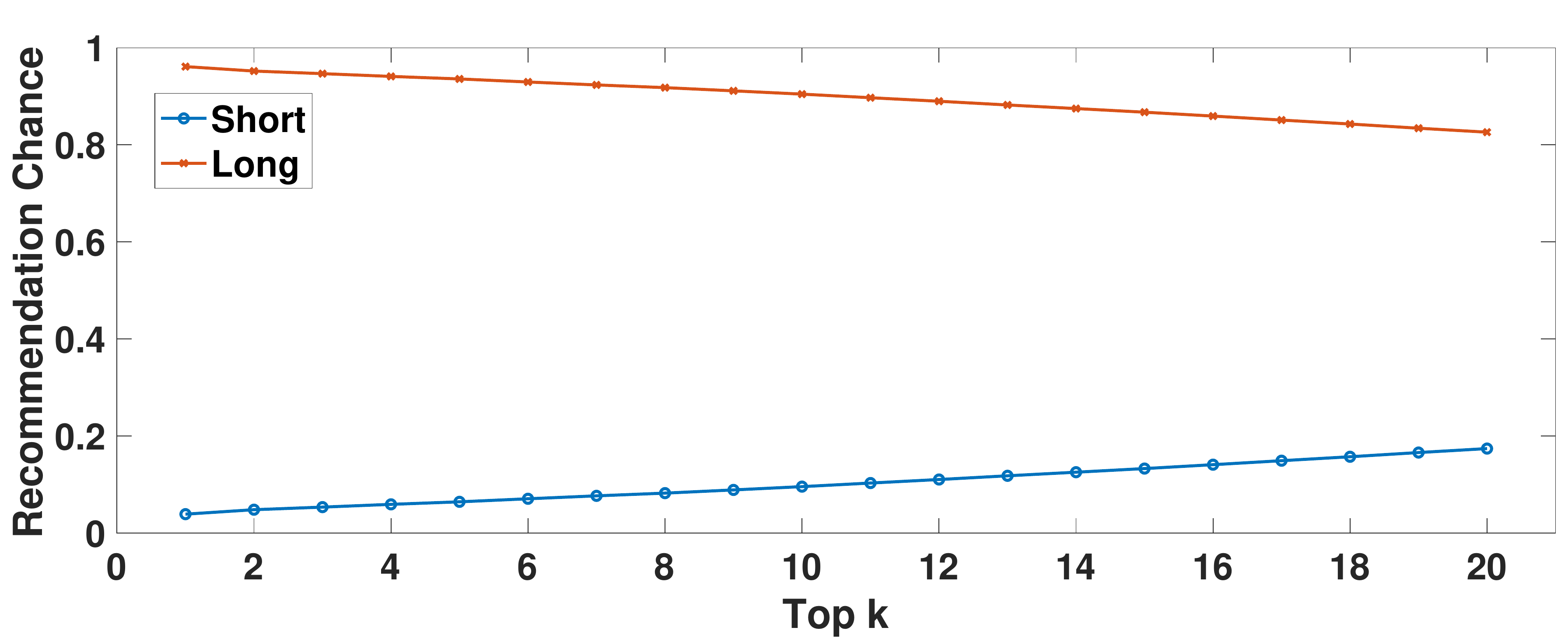}
    \vspace{-15px}
    \caption{Recommendation chances (frequency of being recommended) of two user (producer) groups \textit{w.r.t} top-$k$.}
    \label{fig::traffic_k}
    \vspace{-10px}
\end{figure}

\begin{figure}[t]
    \centering
    \includegraphics[width=\linewidth]{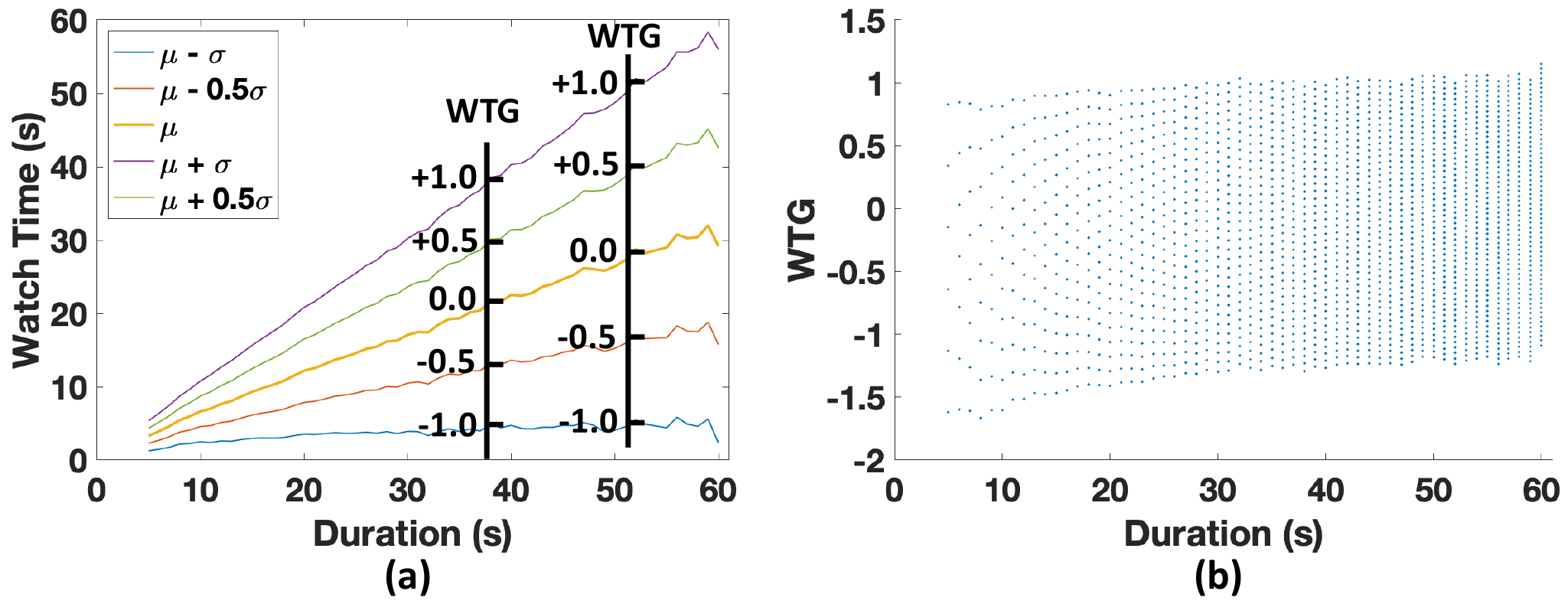}
    \vspace{-15px}
    \caption{(a) Illustration of how WTG is calculated from Watch Time and Video Duration. (b) Distribution of WTG on Wechat Channels dataset. We divide duration into bins with 1 second per bin.}
    \vspace{-15px}
    \label{fig::gain}
\end{figure}

Since videos in the same bin share similar duration, we compare the watch time using data points within each bin, rather than all bins.
Formally, we calculate the mean and standard deviation of watch time in each bin, and then \textbf{WTG} is computed as follows,
\begin{equation}
    \textbf{WTG} = \frac{\textbf{WT} - \mu_{B(v)}}{\sigma_{B(v)}},\label{eq::wtg}
\end{equation}
where \textbf{WT} represents watch time of the data point, $\mu_{B(v)}$ is the mean of watch time in the micro-video $v$'s corresponding bin, and $\sigma_{B(v)}$ is the standard deviation of watch time in that bin.
Both $\mu_{B(v)}$ and $\sigma_{B(v)}$ are calculated from the records of all the users on the whole dataset.
In other words, we standardize the watch time within each bin, which makes WTG independent with micro-video duration.
Intuitively, WTG eliminates the gap between watch time in different bins by normalization, and provides an unbiased measurement of user engagement in micro-videos with different duration.
Figure \ref{fig::gain} (a) illustrates the calculation of WTG, as well as how it is related to watch time and micro-video duration.
By normalizing inside each bin, micro-videos of different duration can be compared by the WTG.
For example, micro-videos of 0 WTG means they are just normal videos regardless of their duration.
And a short video of 1 WTG is definitely better than a long video of 0.5 WTG, even though the original watch time of the long video might be longer.
Figure \ref{fig::gain} (b) shows the distribution of WTG on different micro-video duration, based on the same dataset above.
We can observe that WTG successfully alleviates duration bias since it is distributed more uniformly and does not favor long or short micro-videos.
Moreover, the proposed WTG metric can be efficiently implemented, and the computational details are introduced in Section \ref{app::computation}.

\subsection{DVR: Unbiased Recommendation Model}

\begin{figure}[t]
    \centering
    \includegraphics[width=\linewidth]{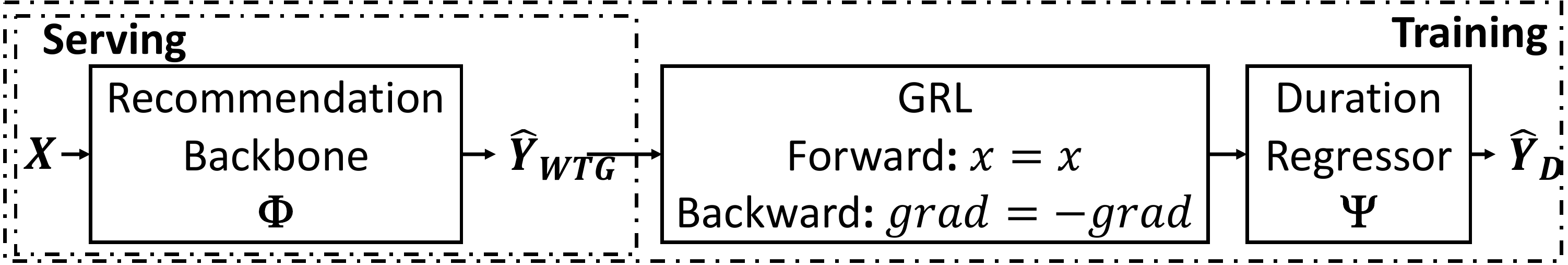}
    \vspace{-15px}
    \caption{Illustration of the proposed DVR framework.}
    \vspace{-15px}
    \label{fig::dvr}
\end{figure}

With the unbiased WTG measurement of user engagement towards micro-videos, we now show how to achieve unbiased recommendation under the guidance of WTG.
Since the structures of backbone models can be quite different, the debiasing design need to be general and compatible for different models.
Therefore, we propose a simple yet effective framework called Debiased Video Recommendation (\textbf{DVR}) which is model-agnostic and it has no preset requirements for backbone models.
Figure \ref{fig::dvr} illustrates the overall design of DVR, where $\Phi$ is the backbone which can be any off-the-shelf recommendation models.
Specifically, we add an adversarial model $\Psi$ on the predicted value of $\Phi$, to make it independent with micro-video duration, thus reduce the impact of duration bias.
We now elaborate on the proposed DVR framework.

\subsubsection{Input Features.}
User profiles and micro-video attributes constitute the input features of recommender systems.
In existing approaches, micro-video duration is included as input features and fed into a machine learning model to predict watch time.
Particularly, it serves as an important feature, to some extent even the most important one, due to the duration bias.
In other words, video duration in the input feature becomes a shortcut for recommendation models to predict watch time directly from it and ignore other features that are related to user preference modeling.
From our above analysis, such duration bias is the key reason of inaccuracy and unfairness.
Therefore, we propose to remove micro-video duration from input features, which eliminates duration bias fundamentally.

\subsubsection{Prediction Target.}
As shown previously in Section \ref{sec::analysis}, recommending directly according to watch time can not well capture user preference and is unfair for short micro-video publishers.
One trivial solution is to transform the predicted watch time to WTG, and then recommend micro-videos according to WTG.
However, it is difficult to predict watch time accurately due to the unbalanced distribution of watch time and duration.
Specifically, the predicted watch time of long (short) videos tend to be longer (shorter), \textit{i.e.} bias amplification shown in Figure \ref{fig::wechat_pred} (a).
Thus, we utilize the proposed unbiased measurement WTG as the prediction target.
During the model training, we optimize recommendation models to predict WTG as accurately as possible.
As for the final recommendation, we rank candidate micro-videos according to the predicted WTG, and then top-$k$ micro-videos with higher ranks are recommended to the user.
Since the distribution of WTG is more uniform and independent with duration as shown in Figure \ref{fig::gain} (b), it is much easier to predict accurately than the biased watch time target.

To evaluate the performance of top-$k$ micro-video recommendation, we further propose \textbf{WTG@}$k$ which is the average of the groundtruth WTG of the top recommended micro-videos as follows,
\begin{equation}
    \textbf{WTG@}k = \frac{1}{k}\sum_{i=1}^k{\mathrm{WTG}(l_i)},
\end{equation}
where $l_i$ is the video with the $i$-th highest predicted WTG.
Thus higher WTG@$k$ means better micro-video recommendation performance, since the user is willing to spend more time watching the recommended micro-videos, compared with the watch time of a random list of micro-videos with a similar duration.

The above WTG@$k$ metric is insensitive to the order of the $k$ recommended micro-videos, which means the same $k$ micro-videos of different orders for a given user will share the same WTG@$k$.
Moreover, in the research field of recommender systems, it is widely acknowledged that higher positions are more important compared with lower ones~\cite{he2017ncf}.
Inspired by the commonly adopted Normalized Discounted Cumulative Gain (NDCG) metric in recommendation \cite{he2017ncf,zheng2021disentangling,chang2021sequential} which emphasize the order of recommended items by assigning larger weights to higher positions, we further propose DCWTG@$k$, which is the discounted cumulative version of WTG.
Specifically, DCWTG@$k$ adds a decaying factor which imposes larger weights on the head of the list, and it is calculated as follows,
\begin{equation}
    \mathrm{DCWTG}@k = \sum_{i=1}^k{\frac{\mathrm{WTG}(l_i)}{\mathrm{log}_2(1 + i)}}.
\end{equation}

\subsubsection{Model Training with Adversarial Learning}
We now investigate how to capture unbiased user preference that is free from the influence of micro-video duration.
Although we remove duration from input features and use the unbiased WTG as the prediction target, the influence of duration bias can not be fully eliminated and it still hides implicitly in the data, \textit{e.g.} duration can be correlated with other input features like micro-video category.
Therefore, in order to make the predicted WTG independent with micro-video duration, we add an extra regression layer, denoted as $\mathbf{\Psi}$, to predict duration from the estimated WTG and train the recommendation model, denoted as $\mathbf{\Phi}$, in an adversarial way.
Specifically, we encourage the extra regression layer to predict micro-video duration as accurately as possible, and force the recommendation model to best fool the regression layer.
In short, it follows a manner of adversarial learning, which can be formally denoted as follows,

\begin{align}
    &\hat{Y}_{WTG} = \mathbf{\Phi}(X), \label{eq::y_wtg}\\
    &\hat{Y}_{D} = \mathbf{\Psi}(\hat{Y}_{WTG}), \label{eq::y_d}
\end{align}
where X is the input features, $\hat{Y}_{WTG}$ and $\hat{Y}_{D}$ are the predicted WTG and duration, respectively.

As for duration regression model $\mathbf{\Psi}$, we force it to discover possible correlations between the predicted WTG from $\mathbf{\Phi}$ and micro-video duration as much as possible.
As for the recommendation model $\mathbf{\Phi}$, the adversarial learning encourages it to squeeze out all the information about micro-video duration.
In other words, by adding an extra regression model $\mathbf{\Psi}$, the recommendation model $\mathbf{\Phi}$ learns to predict WTG without being disturbed by micro-video duration.
Inspired by the recent advances\cite{ganin2015unsupervised,zheng2021dgcn}, 
we implement the adversarial learning by inserting a Gradient Reversal Layer (GRL)  between $\mathbf{\Phi}$ and $\mathbf{\Psi}$, as illustrated in Figure \ref{fig::dvr}.
In this way, the recommendation model $\mathbf{\Phi}$ captures unbiased user preference, which is free from the notorious duration bias.
Then we have two loss functions for regression as follows,
\begin{align}
    &L_{WTG} = \textbf{MSE}(\hat{Y}_{WTG}, Y_{WTG}), \label{eq::l_wtg}\\
    &L_{D} = \textbf{MSE}(\hat{Y}_{D}, Y_{D}), \label{eq::l_d}
\end{align}
where $Y_{WTG}$ and $Y_{D}$ are the groundtruth value of WTG and duration.
Here \textbf{MSE} represents the \textit{Mean Squared Loss} function.
To balance the two loss functions, we add a hyper-parameter $\alpha$, which controls the intensity of adversarial learning.
The two components, $\mathbf{\Phi}$ and $\mathbf{\Psi}$, are optimized with $L_{WTG}$ and $L_{D}$ in an end-to-end manner.
We show the whole process of DVR in Algorithm \ref{alg::dvr}.

\subsubsection{Discussion of Backbone Recommendation Model}
It is worthwhile to notice that the proposed DVR approach is highly general and can be integrated with any off-the-shelf recommendation models,
since we impose no restrictions on the structure of $\mathbf{\Phi}$.
Specifically, the extra duration regression component $\mathbf{\Psi}$ can be appended on any appropriate $\mathbf{\Phi}$ that can perform real-value regression from high-dimensional input features.
For example, existing deep learning based recommendation models \cite{guo2017deepfm,he2017nfm,song2019autoint,cheng2020adaptive} are perfect candidates for $\mathbf{\Phi}$.
We will show in experiments (Section \ref{sec::exp}) that DVR can achieve consistent improvements in both fairness and accuracy when combined with different backbone recommendation models.

\noindent\textbf{Remark.}
In real-world micro-video applications, multi-task learning framework~\cite{gao2019neural,jin2020multi} is usually adopted, which includes targets other than watch time, such as like, comment, follow, and so on.
Although these signals may not be affected by the duration bias, they are hard to collect (very sparse in the real world), while watch time is the most fundamental user feedback in micro-video platforms~\cite{covington2016deep}. Therefore, our solution is essential and practical.

\begin{algorithm}[tb]
\caption{Debiased Video Recommendation (DVR)}
\label{alg::dvr}
\begin{flushleft}
\textbf{Input}: Training data $\mathcal{O}$ with features $X$, watch time labels $Y_{WT}$, and duration labels $Y_{D}$\\
\textbf{Models}: WTG regression model $\mathbf{\Phi}$, duration regression model $\mathbf{\Psi}$\\
\end{flushleft}
\begin{algorithmic}[1] %
\STATE remove duration from features $X$
\STATE compute WTG labels $Y_{WTG}$ by Algorithm \ref{alg::code}/\ref{alg::online_wtg}
\WHILE{not converge}
\FOR{batch in $\mathcal{O}$}
\STATE compute $\hat{Y}_{WTG}$, $\hat{Y}_{D}$ according to (\ref{eq::y_wtg})-(\ref{eq::y_d})
\STATE compute $L_{WTG}$, $L_{D}$ according to (\ref{eq::l_wtg})-(\ref{eq::l_d})
\STATE optimize $\mathbf{\Psi}$ with $\alpha L_{D}$
\STATE optimize $\mathbf{\Phi}$ with $L_{WTG}~-~\alpha L_{D}$
\ENDFOR
\ENDWHILE
\end{algorithmic}
\end{algorithm}

%% file: 4.experiments.tex
\section{Experiments}\label{sec::exp}

\subsection{Experimental Settings}

\subsubsection{\textbf{Datasets.}}\label{sec::dataset}
We utilize two public real-world datasets, which are collected from two large micro-video platforms, Wechat Channels\footnote{\url{https://www.wechat.com/en}} and Kuaishou\footnote{\url{https://www.kuaishou.com}}.
Each dataset is composed of abundant micro-videos of different duration, and both exhibits strong duration bias, \textit{e.g.} duration bias of Wechat dataset has been shown in Figure \ref{fig::duration_bias}.
The details of the adopted datasets are introduced in Section \ref{app::dataset}.

\subsubsection{\textbf{Backbone Models.}}
To investigate the recommendation performance, we experiment with both classical and state-of-the-art recommendation backbone models, including \textbf{LibFM} \cite{rendle2010factorization}, \textbf{Wide\&Deep} \cite{cheng2016wide}, \textbf{DeepFM} \cite{guo2017deepfm}, \textbf{NFM} \cite{he2017nfm}, \textbf{AFM} \cite{xiao2017attentional}, \textbf{AutoInt} \cite{song2019autoint} and \textbf{AFN} \cite{cheng2020adaptive}.
Details of all the models are in Section \ref{app::baseline}.

\subsubsection{\textbf{Metrics.}}
To evaluate the performance of learning user preference, we calculate the two proposed metrics, WTG@$k$ and DCWTG@$k$.
Both metrics are calculated for each user, and we report the average value of all users.
Higher WTG@$k$ and DCWTG@$k$ mean better recommendation performance.
We also evaluate the number of bad cases (\#BC@$k$) for each model, which is defined previously in Table \ref{tab::analysis} as the number of recommended videos with watch time less than 2 seconds.
It is worthwhile to note that lower \#BC@$k$ means better recommendation performance.
$k$ is set as 10 in our experiments, a widely selected value~\cite{he2017ncf}, which measures the quality of the top 10 recommended videos.

Implementation details are introduced in Section \ref{app::implementation}.

\subsection{Effectiveness of WTG (RQ1)}

For each recommendation model, we train two versions of it, using watch time or WTG as the target, respectively.
We stop training when the regression accuracy on the validation set converges.
During the evaluation, for each user, the two versions predict watch time or WTG of the micro-videos in the test set, then micro-videos are ranked according to the estimated watch time or WTG, respectively.
Finally, top $k$ micro-videos with the highest estimated watch time or WTG are recommended to each user.
We compare the two versions with respect to both accuracy and fairness.

\begin{table*}[t!]
\centering
\small
\caption{Recommendation performance comparison of different backbone models with/without DVR on two datasets.}
\label{tab::overall}
\begin{tabular}{cc|ccc|ccc}
\toprule
\multicolumn{2}{c|}{\textbf{Method}}          & \multicolumn{3}{c|}{\textbf{Wechat}} & \multicolumn{3}{c}{\textbf{Kuaishou}} \\
Backbone                    & \textbf{Debias} & WTG@10    & DCWTG@10    & \#BC@10    & WTG@10     & DCWTG@10    & \#BC@10    \\
\midrule
\multirow{3}{*}{FM}         & None            & 0.0209          & 0.2985          & 6381          & 0.0571          & 0.4178          & 5854          \\
                            & DVR-            & 0.1249          & 1.3813          & 5994          & 0.1662          & 1.1318          & 5728          \\
                            & DVR             & \textbf{0.1332} & \textbf{1.5100} & \textbf{5947} & \textbf{0.2094} & \textbf{1.6137} & \textbf{5240} \\ \hline
\multirow{3}{*}{WDL}        & None            & 0.0265          & 0.3880          & 6326          & 0.0532          & 0.4031          & 5851          \\
                            & DVR-            & 0.1342          & 1.4683          & 5926          & 0.2002          & 1.4810          & 5511          \\
                            & DVR             & \textbf{0.1468} & \textbf{1.6539} & \textbf{5881} & \textbf{0.2087} & \textbf{1.5833} & \textbf{5226} \\ \hline
\multirow{3}{*}{DeepFM}     & None            & 0.0236          & 0.3648          & 6345          & 0.0550          & 0.4161          & 5843          \\
                            & DVR-            & 0.1372          & 1.5086          & 5894          & \textbf{0.2132} & 1.5664          & 5426          \\
                            & DVR             & \textbf{0.1469} & \textbf{1.6551} & \textbf{5866} & 0.2066          & \textbf{1.5902} & \textbf{5261} \\ \hline
\multirow{3}{*}{NFM}        & None            & 0.0234          & 0.3334          & 6345          & 0.0561          & 0.4478          & 5826          \\
                            & DVR-            & 0.1302          & 1.4338          & 5952          & \textbf{0.2089} & 1.5632          & 5368          \\
                            & DVR             & \textbf{0.1444} & \textbf{1.6226} & \textbf{5899} & 0.2081          & \textbf{1.6050} & \textbf{5230} \\ \hline
\multirow{3}{*}{AFM} & None            & 0.0335          & 0.4028          & 6349          & 0.1052          & 0.7237          & 6337          \\
                            & DVR-            & 0.1203          & 1.3318          & 5986          & 0.1260          & 0.8890          & 5726          \\
                            & DVR             & \textbf{0.1391} & \textbf{1.5656} & \textbf{5930} & \textbf{0.2082} & \textbf{1.6068} & \textbf{5209} \\ \hline
\multirow{3}{*}{AutoInt}    & None            & 0.0272          & 0.3862          & 6330          & 0.0504          & 0.3823          & 5868          \\
                            & DVR-            & 0.1351          & 1.4841          & 5924          & \textbf{0.2124} & 1.5561          & 5343          \\
                            & DVR             & \textbf{0.1458} & \textbf{1.6420} & \textbf{5874} & 0.2086          & \textbf{1.5905} & \textbf{5237} \\ \hline
\multirow{3}{*}{AFN}        & None            & 0.0157          & 0.2599          & 6358         & 0.0536          & 0.4037          & 5832          \\
                            & DVR-            & 0.1254          & 1.3714          & 6064          & 0.1691          & 1.2442          & 5552          \\
                            & DVR             & \textbf{0.1408} & \textbf{1.5858} & \textbf{5917} & \textbf{0.2015} & \textbf{1.5551} & \textbf{5229} \\ 
\bottomrule
\end{tabular}
\end{table*}

\begin{figure}[t]
    \centering
    \includegraphics[width=\linewidth]{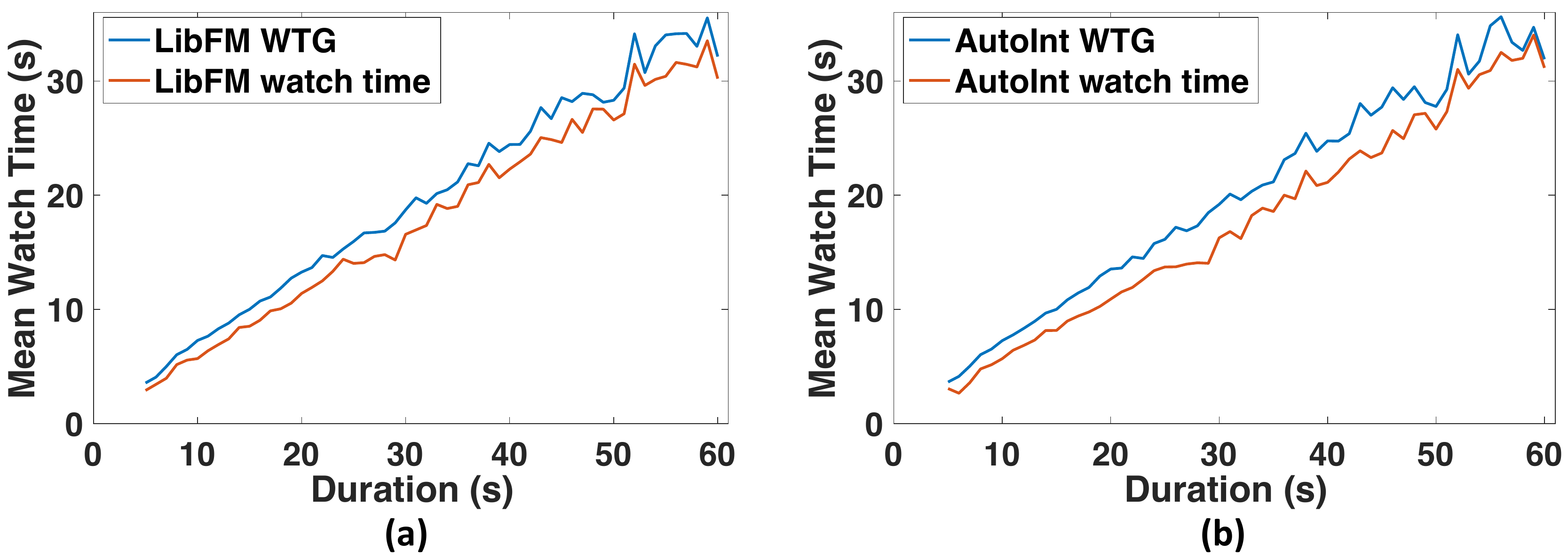}
    \vspace{-15px}
    \caption{Accuracy comparison between Watch Time and WTG. We plot the mean watch time of recommended videos. Two selected models: (a) LibFM (b) AutoInt.}
    \vspace{-10px}
    \label{fig::rec_quality}
\end{figure}

\begin{figure}[t]
    \centering
    \includegraphics[width=\linewidth]{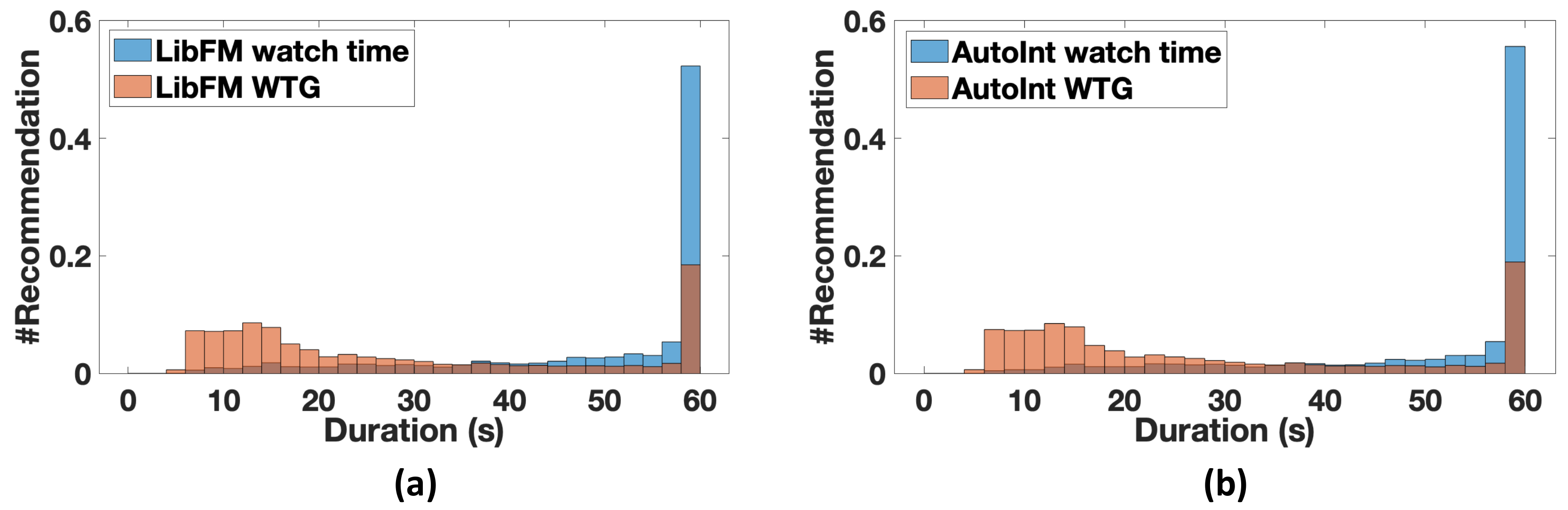}
    \vspace{-15px}
    \caption{Fairness comparison between Watch Time and WTG. We plot the histogram on recommended videos of different duration. Two selected models: (a) LibFM (b) AutoInt.}
    \vspace{-15px}
    \label{fig::metric_comparison}
\end{figure}

\noindent \textbf{Accuracy Comparison of Watch Time v.s. WTG.}
Recommendation according to watch time may lead to low accuracy since there are many bad cases where the user only watches a few seconds of a long micro-video.
To illustrate this point, we investigate the accuracy of recommended micro-videos of different duration.
Specifically, we separate the recommended micro-videos into bins according to their duration, and then calculate the average ground-truth watch time of the recommended micro-videos.
Figure \ref{fig::rec_quality} shows the results of recommending according to watch time or WTG.
We can observe that models trained with watch time as target result in inferior recommendation quality, since the average ground-truth watch time of recommended micro-videos is consistently lower than trained with WTG.
With the unbiased WTG as the prediction target, models can generate much more high-quality recommendations of both long and short micro-videos. 

\noindent \textbf{Fairness Comparison of Watch Time v.s. WTG.}
Figure \ref{fig::metric_comparison} illustrates the histogram on the duration of recommended micro-videos for LibFM and AutoInt.
Results of the other five backbones are similar and we omitted due to the space limitation.
We can observe that the recommended micro-videos from models trained with watch time as target mainly concentrate on the long duration side, which is because watch time is to a great extent dominated by the duration bias.
For example, videos of 60s (the maximum length in Wechat dataset) take over 52.3\% and 55.6\% of recommendation traffic for LibFM and AutoInt, respectively, while short videos of less than 20s almost receive no recommendation with recommendation traffic lower than 1.46\%.
Using WTG as the target can largely solve this problem, and we can discover that the duration distribution of recommended micro-videos from models trained with WTG as the target is much more balanced compared with using watch time as the target.
Specifically, videos of 60s only receive about 18.47\% and 18.95\% of recommendation for LibFM and AutoInt, which is far less than using watch time as the target, and short videos of less than 20s also receive fair enough recommendation chance with about 8.59\% of total recommendation.
In other words, the proposed WTG serves as an unbiased target to train recommendation models, and achieves fair recommendation of micro-videos with different duration, which does not favor long or short micro-video publishers.

In summary, the proposed unbiased target WTG successfully improves both the accuracy and fairness of recommendation models against the biased watch time target.

\subsection{Effectiveness of DVR (RQ2)}

We combine DVR with all the recommendation backbones, and Table \ref{tab::overall} shows the results.
We also include a simplified version of the proposed model called \textbf{DVR-}, which means that the backbone model is trained with watch time as target while we transform the predicted watch time to the proposed metric WTG for ranking.
We have the following observations:
\begin{itemize}[leftmargin=*]
    \item \textbf{Worse performance of existing recommendation models.}
    Without special designs to eliminate duration bias, existing recommendation approaches can not well capture user preference, and they are easily misled to blindly recommend micro-videos with long duration.
    However, these long micro-videos fail to meet users' interest, and users only watch a few seconds of them, leading to low WTG and DCWTG.
    Note that in equation (\ref{eq::wtg}) WTG is a normalized metric by subtracting mean watch time then divided by the standard deviation, thus WTG close to zero means that the recommended micro-videos are almost as the same quality as random recommendation.
    We can observe that the WTG of top-10 micro-videos is very low for all recommendation models without debiasing, which verifies that duration bias leads to bad recommendation accuracy.
    Meanwhile, \#BC@10 of backbone models without any debiasing design is much higher than DVR- and DVR, which means that recommending according to watch time provides a large amount of unsatisfactory micro-videos, which may directly lead to user churn.
    \item \textbf{Steady improvement of our DVR model.} The proposed DVR can improve recommendation accuracy significantly.
    Specifically, the progress of WTG@10 is over 300\% in most cases of seven backbones on two datasets.
    For the state-of-the-art method AFN, DVR can improve WTG by over 500\%.
    In addition, the number of bad cases for DVR is significantly less than simply using the backbone models.
    For example, \#BC@10 of DVR is about 21.65\% less than AFM on Kuaishou dataset.
    Meanwhile, consistent improvements across different backbones demonstrate that DVR is a highly general framework that can be integrated smoothly with existing recommendation approaches.
    Another interesting finding is that although DVR- is worse than DVR in most cases, it outperforms backbone models with significant improvements.
    In fact, DVR- utilizes well-trained biased models, and corrects the duration bias directly from the predicted watch time, by transforming it to the unbiased WTG value.
    The huge improvements of DVR- over backbone models indicate that it is easy to apply our proposed WTG to existing recommendation systems.
\end{itemize}

\textit{Due to space limit, more experimental results including ablation study and hyper-parameter study of DVR can be found in Section \ref{app::exp}.}

%% file: 5.related.tex
\vspace{-0.1cm}
\section{Related Work}\label{sec::related}
\noindent{\textbf{Video Recommendation.}}
Users are spending more and more time in video apps, especially micro-video apps such as TikTok and Kuaishou.
As the number of uploaded videos is quite large, it is critical to utilize a recommender system to provide personalized videos to users \cite{davidson2010youtube,covington2016deep,beutel2018latent,chen2019top,wei2019mmgcn,liu2019user,jiang2020aspect,li2019routing,chen2018temporal,gao2017unified,wei2019personalized,yu2020deep}.
For example, the YouTube recommendation has evolved from rule-based systems \cite{davidson2010youtube}, to Deep Neural Networks (DNN) based models \cite{covington2016deep}, then Recurrent Neural Networks (RNN) based models \cite{beutel2018latent}, and now Reinforcement Learning (RL) based models \cite{chen2019top}.
Li \textit{et al.} \cite{li2019routing} proposed to capture user interest by leveraging multiple user behaviors towards micro-videos such as click, like and follow, with a graph-based Long Short-Term Memory (LSTM) model.
In addition, Wei \textit{et al.} \cite{wei2019mmgcn} proposed a Graph Convolutional Networks (GCN) based model which leverages multi-modal information to enhance the performance of short video recommendation.
However, these approaches either focus on traditional discrete user feedback like clicks or predict the continuous watch time feedback, which exhibits strong bias.
To the best of our knowledge, we are the first to reduce duration bias for micro-video recommendation, which is crucial for learning users' real interests that are independent of video duration.

\noindent\textbf{Fairness-aware Recommendation.}
As recommender systems grow increasing impact on users, fairness becomes a critical issue \cite{beutel2019fairness,fu2020fairness,li2021towards,ge2022toward,li2021tutorial,wu2021tfrom,zhu2021fairness,wadsworth2018achieving,beutel2017data,zhang2018mitigating,adel2019one}, especially in user-generated content (UGC) platforms where multi-stakeholders are involved, such as micro-video applications.
Fairness-aware recommendation is generally studied from two perspectives \cite{li2021tutorial,mehrotra2018towards}, including user fairness which focuses on algorithmic bias towards specific individuals or user groups \cite{leonhardt2018user,li2021user}, and item fairness which means fair recommendation traffic received by different items \cite{abdollahpouri2017controlling,li2021tutorial,mehrotra2018towards}.
Unlike existing fairness-aware recommendation literature, in this paper, we address a specific fairness issue in micro-video platforms, where micro-videos
of different duration tend to receive unfair recommendation traffic.

\noindent{\textbf{Duration Bias.}}
Bias in recommender systems \cite{chen2020bias} has been studied from several directions, such as popularity bias \cite{gilotte2018offline,zheng2021disentangling,zhang2021causal} and position bias \cite{collins2018study,o2006modeling}.
However, duration bias in video recommendation has been unexplored until a recent study \cite{wu2018beyond}, in which Wu \textit{et al.} investigated the bias of watch time and watch percentage from an aggregated level, \textit{i.e.} the average of the watch time of all users towards each video.
In other words, it merges all samples of the same video into one single data point, and compares with other videos to measure the video quality.
Unlike~\cite{wu2018beyond}, our study focuses on the personalized duration bias, where different users have distinct WTG values towards the same micro-video.
Our setting is closer to the real-world recommendation scenarios, and the proposed WTG metric can be directly integrated into online recommender systems.

%% file: 6.conclusion.tex
\section{Conclusion and Future Work}\label{sec::conclusion}
In this paper, we investigate a largely unexplored duration-bias problem in micro-video recommendation.
We conduct large-scale data analysis to show that the duration bias leads to inaccurate and unfair recommendation.
A new measurement of watch time on micro-videos, WTG, is proposed which eliminates duration bias and can evaluate recommendation performance without favoring either long or short videos.
A general model DVR is further designed to help recommendation models learn unbiased user preferences.
Experiments demonstrate that the proposed metric and model successfully eliminate duration bias, which can achieve accurate and fair recommendation.
As for the future work, we plan to apply WTG in online systems to evaluate the performance of micro-video recommendation.
We also plan to evaluate DVR with online A/B tests.

%% file: 7.appendix.tex
\section{Appendix}

\subsection{Offline and Online Computation of WTG.}\label{app::computation}
With respect to offline evaluation, WTG can be efficiently computed from the entire data of recommender systems.
We provide the pseudocode for calculating WTG in Algorithm \ref{alg::code}.

However, in terms of online serving, what recommender systems handle are streams of unstopped logs.
Thus WTG is dynamically changing since the arriving new data from the stream influences the mean and standard deviation of watch time in each bin.
Fortunately, the mean and standard deviation can be updated in a recursive manner with no need to save all the data \cite{leskovec2020mining}.
Specifically, we only need three extra variables ($\mu$, $\sigma$, and $n$) to keep track of the mean and standard deviation of watch time, as well as the number of data points in each bin.
These three variables are updated dynamically according to the data stream, and Algorithm \ref{alg::online_wtg} briefly illustrates such process.
It is worthwhile to notice that our provided Algorithm \ref{alg::online_wtg} is just a sketch for the online serving of WTG, and there can be more efficient implementations. Nevertheless, our purpose is to show that the proposed metric can be seamlessly integrated into the online recommendation systems in a real-time streaming manner.

\begin{algorithm}[h]
\caption{Offline Computation of WTG}
\label{alg::code}
\begin{flushleft}
\textbf{Input}: Dataframe $D$ of format $(user, video, watchtime, duration)$\\
\end{flushleft}
\begin{algorithmic}[1] %
\STATE{$DG \leftarrow \mathrm{GroupBy}(D.duration)$}
\STATE{$D_{mean} \leftarrow \mathrm{Mean}(DG.watchtime)$}
\STATE{$D_{std} \leftarrow \mathrm{Std}(DG.watchtime)$}
\STATE{$D.mean, D.std \leftarrow \mathrm{Join}(D, D_{mean}, D_{std})$}
\STATE{$D.wtg \leftarrow (D.watchtime - D.mean)/D.std$}
\end{algorithmic}
\end{algorithm}

\begin{algorithm}[h]
\caption{Online Computation of WTG}
\label{alg::online_wtg}
\begin{flushleft}
\textbf{Input}: Data stream of application logs $S$\\
\textbf{Tracking Variables}: Mean of watch time $[\mu_1, \cdots, \mu_m]$, std of watch time $[\sigma_1, \cdots, \sigma_m]$, and the number of data points $[n_1, \cdots, n_m]$ in $m$ different bins
\end{flushleft}
\begin{algorithmic}[1] %
\WHILE{$S$ is not empty}
\STATE{$(\mathrm{WT}, d_v) \leftarrow S.pop()$ ~// Get watch time and duration.}
\STATE{$b \leftarrow f_b(d_v)$ ~~// Get the corresponding bin.}
\STATE{$n_b \leftarrow n_b + 1$ ~~// Update the number of data points.}
\STATE{$\sigma^2_b \leftarrow \frac{n_b - 1}{n^2_b}(\mathrm{WT} - \mu_b)^2 + \frac{n_b - 1}{n_b}\sigma^2_b$ ~// Update std.}
\STATE{$\mu_b \leftarrow \mu_b + \frac{\mathrm{WT} - \mu_b}{n_b}$ ~// Update mean.}
\ENDWHILE
\end{algorithmic}
\end{algorithm}

\subsection{Details of the Adopted Datasets}\label{app::dataset}

We utilize two public real-world datasets, both of which are collected from large short-video platforms.
We summarize the statistics of the adopted datasets in Table \ref{tab::dataset}, where we also list the total duration (in seconds) of all records.
The details of the two adopted datasets are as follows,
\begin{itemize}[leftmargin=*]
    \item \textbf{Wechat}: This dataset is released by WeChat Big Data Challenge 2021\footnote{\url{https://algo.weixin.qq.com/}} which contains the logs on Wechat Channels within two weeks.
    We split the data into the first ten days, the middle two days, and the last two days as training, validation, and test set.
    The adopted input features include \textit{UserID}, \textit{VideoID}, \textit{DeviceID}, \textit{AuthorID}, \textit{BGMSongID}, \textit{BGMSingerID}, \textit{UserActiveness}, and \textit{VideoPopularity}.
    \item \textbf{Kuaishou}: This dataset \cite{li2019routing} is released by the Kuaishou Competition in China MM 2018 Conference\footnote{\url{https://github.com/liyongqi67/ALPINE}}, and we also split the datasets into training, validation, and test sets according to timestamps with the splitting ratio as 8:1:1. The adopted input features include \textit{UserID}, \textit{VideoID}, \textit{UserActiveness}, and \textit{VideoPopularity}.
\end{itemize}

Since we divide all the micro-videos into separate bins according to their duration and compute the mean and standard deviation of watch time, each bin is supposed to have enough data points to guarantee that the computed $\mu$ and $\sigma$ are statistically significant.
Therefore, we filter out those bins of too long or too short duration, which only contain a few data points.
Specifically, for the Wechat dataset, we reserve the micro-videos with a duration between 5 seconds and 60 seconds, and for the Kuaishou dataset, we keep the micro-videos longer than 5 seconds and shorter than 120 seconds.
Micro-videos with duration outside the above range are of low prevalence and they only take less than 0.1\% of all the records.
After filtering out too long or too short micro-videos, for both datasets, we construct equally wide bins with 1 second per bin.
It is worthwhile to note that each bin contains over 10,000 data points which guarantees the statistical significance of the computed mean and standard deviation values.

\begin{table}
    \caption{Statistics of two adopted real-world datasets.}
    \vspace{-0.4cm}
    \label{tab::dataset}
    \begin{tabular}{ccccc}
      \toprule
      \bf{Dataset} & \bf{\#Users} & \bf{\#Videos} & \bf{\#Records} & \bf{Total Duration (s)}  \\
      \midrule
      Wechat & 10,000 & 639,557 & 2,672,809 & 46,785,442  \\
      Kuaishou & 20,000 & 96,418 & 7,310,108 & 227,955,046 \\
      \bottomrule
    \end{tabular}
\end{table}

\subsection{Details of Backbone Models}\label{app::baseline}
We include the following recommendation backbone models,
\begin{itemize}[leftmargin=*]
    \item \textbf{LibFM} \cite{rendle2010factorization}. This is a classical recommendation algorithm which captures feature interaction by taking inner product of each pair of features.
    \item \textbf{Wide\&Deep} \cite{cheng2016wide}. It combines linear regression and deep neural networks to learn direct feature matching and high-order feature interaction separately.
    \item \textbf{DeepFM} \cite{guo2017deepfm}. This method ensembles multi-layer perceptions (MLP) and LibFM.
    \item \textbf{NFM} \cite{he2017nfm}. This method extends LibFM with a Bi-Interaction layer.
    \item \textbf{AFM} \cite{xiao2017attentional}. It utilizes attention to aggregate different cross features in LibFM.
    \item \textbf{AutoInt} \cite{song2019autoint}. It utilizes multi-head self-attention to automatically construct complex feature interactions.
    \item \textbf{AFN} \cite{cheng2020adaptive}. This is the state-of-the-art method which learns arbitrary order of feature interaction with a logarithmic transformation layer.
\end{itemize}

\subsection{Implementation Details}\label{app::implementation}
We implement all the backbone models, and the proposed DVR with TensorFlow \cite{abadi2016tensorflow}.
We use Adam \cite{kingma2014adam} as the optimizer and set the initial learning rate as 0.001.
The batch size is set as 512.
For a fair comparison, we use three hidden layers and 64 hidden units per layer for all models using DNN.
We train the models until convergence and use early stopping to avoid overfitting.
For DVR, $\Psi$ is implemented as a $1 \times 1$ dense layer, and the optimal $\alpha$ is 0.1.
Other hyper-parameters of all these models are tuned carefully on the validation set using grid search, following settings or suggestions of original papers.
We have released the code and data at  \url{https://github.com/tsinghua-fib-lab/WTG-DVR}.

\subsection{More Experimental Results}\label{app::exp}

\subsubsection{Ablation studies of our DVR model.}
There are three key strategies in DVR, which are DD (delete duration from input features), WTG (use WTG as the target instead of watch time), and ADV (Adversarial training).
We investigate the contribution of each component by adding the three strategies one by one.
Table \ref{tab::ablation} shows the DCWTG@10 of two typical cases. The results of other cases are similar and omitted due to space limitation.
We can observe that simply deleting micro-video duration from input features can bring about 10\% and 50\% improvements on two datasets, respectively.
Meanwhile, the largest improvements come from introducing the proposed unbiased WTG as the prediction target.
Moreover, adversarial learning can further improve the recommendation performance by about 10\%.
In fact, the three key strategies eliminate duration bias from three different perspectives, which are input, output, and model itself.
Combining the three simple yet effective strategies leads to fair and accurate recommendation of micro-videos.

\subsubsection{Hyper-parameter study of DVR.}
\begin{figure}[t]
    \centering
    \includegraphics[width=\linewidth]{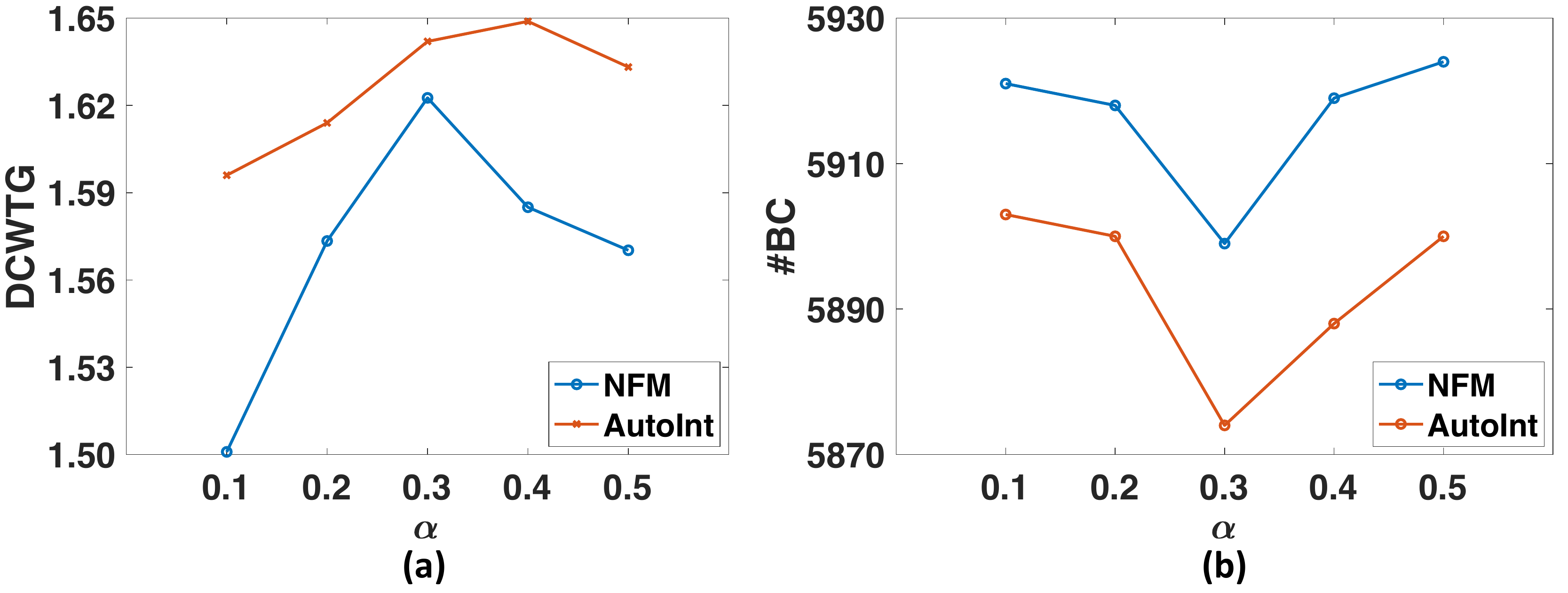}
    \caption{Performance of NFM and AutoInt on Wechat dataset under different values of $\alpha$ with respect to (a) DCWTG (b) Number of bad cases.}
    \label{fig::hyper_param}
\end{figure}

In the proposed DVR model, we introduce a hyper-parameter $\alpha$, the loss weight, which controls the intensity of adversarial learning.
Figure \ref{fig::hyper_param} illustrates the recommendation performance of NFM and AutoInt under different values of $\alpha$.
We can observe that setting $\alpha$ as 0.3-0.4 achieves the best performance with respect to both DCWTG and \#BC. 
On the one hand, low $\alpha$ such as 0.1 imposes too weak adversarial supervision on the recommendation model.
In other words, the duration regressor $\Psi$ receives insufficient optimization, which can not provide much help for the recommendation backbone model $\Phi$.
As a consequence, the prediction of $\Phi$ is still correlated with video duration, which leads to less gain and more bad cases.
On the other hand, if $\alpha$ is too high, the auxiliary adversarial task may interfere with the main task.
Specifically, the adversarial signals from the duration regressor $\Psi$ becomes dominant of the optimization, which makes the recommendation backbone $\Phi$ prone to underfitting.

\begin{table}[t]
\centering
\caption{Ablation study of DVR.}
\label{tab::ablation}
\begin{tabular}{cccccc}
\toprule
\bf{Dataset}  & \bf{Model} & \bf{None} & \bf{+DD}      & \bf{+WTG}      & \bf{+ADV}      \\
\midrule
Wechat   & NFM      & 0.3334    & +8.59\%  & +354.28\% & +386.68\% \\
Kuaishou & AFN      & 0.4037   & +49.08\% & +267.39\% & +285.21\% \\
\bottomrule
\end{tabular}
\end{table}

